\documentclass[conference]{IEEEtran}
\IEEEoverridecommandlockouts
\usepackage{cite}
\usepackage{amsmath,amssymb,amsfonts}
\usepackage{wasysym}
\usepackage{siunitx}
\usepackage{algorithmic}
\usepackage{graphicx}
\usepackage{subcaption}
\usepackage{flafter}
\usepackage{placeins}[section]

\usepackage{textcomp}
\usepackage{xcolor}
\usepackage{mathtools}
\def\BibTeX{{\rm B\kern-.05em{\sc i\kern-.025em b}\kern-.08em
    T\kern-.1667em\lower.7ex\hbox{E}\kern-.125emX}}
\usepackage{hyperref}
\usepackage[noabbrev,capitalise]{cleveref}
\Crefname{figure}{Fig.}{Figs.}
\usepackage{tcolorbox}

\title{Coordination of Heterogeneous Deferrable Loads using the F-MBC Mechanism}

\author{\IEEEauthorblockN{Subhitcha Ramkumar}
\IEEEauthorblockA{\textit{Dept.~of Electrical Sustainable Energy}\\
\textit{Delft University of Technology}\\
Delft, The Netherlands \\
subhitcharamkumar@outlook.com}
\and
\IEEEauthorblockN{ Hazem A. Abdelghany\IEEEauthorrefmark{1}}\thanks{\IEEEauthorrefmark{1} HAA has received funding from the European Union’s Horizon 2020 research and innovation programme under Marie Sklodowska-Curie grant agreement No. 675318 (INCITE). HAA is also with the electrical and control engineering department, Arab Academy for Science, Technology and Maritime Transport, Cairo, Egypt.}
\IEEEauthorblockA{
\textit{Dept.~of Electrical Sustainable Energy}\\
\textit{Delft University of Technology}\\
Delft, The Netherlands \\
h.a.m.f.abdelghany@tudelft.nl}
\and
\IEEEauthorblockN{Simon H. Tindemans}
\IEEEauthorblockA{\textit{Dept.~of Electrical Sustainable Energy}\\
\textit{Delft University of Technology}\\
Delft, The Netherlands\\
s.h.tindemans@tudelft.nl}
}

\begin{document}

\IEEEoverridecommandlockouts
\IEEEpubid{\parbox{\columnwidth}{\copyright 2021 IEEE. Personal use of this material is permitted. Permission from IEEE must be obtained for all other uses, in any current or future media, including reprinting/republishing this material for advertising or promotional purposes, creating new collective works, for resale or redistribution to servers or lists, or reuse of any copyrighted component of this work in other works.}\hspace{\columnsep}\makebox[\columnwidth]{ }}

\maketitle

\IEEEpubidadjcol

\begin{abstract}
Increasing participation of prosumers in the electricity grid calls for efficient operational strategies for utilizing the flexibility offered by Distributed Energy Resources (DER) to match supply and demand. This paper investigates the coordination performance of a recently proposed coordination scheme for deferrable loads: Forecast Mediated Market Based Control (F-MBC). Enhancements are made to the simulation setup to enable an analysis of performance in realistic scenarios, with heterogeneous loads and an open-ended simulation horizon. Operational scenarios were formulated to showcase the ability of F-MBC to schedule heterogeneous populations of deferrable loads with dynamic load profiles, supported by a mix of renewable and flexible generation. Availability patterns of devices were generated to take into account varying user preferences. Simulation results indicate that F-MBC was able to achieve good distributed scheduling performance for devices with a high initial power consumption. However, performance for devices with low initial power consumption has been found to be less satisfactory. Several directions for further improvement of the F-MBC scheme and its applications are identified.
\end{abstract}

\section{Introduction}
 The ongoing shift towards smart grids from the conventional centralized nature of electricity generation has led to an increased proliferation of Distributed Energy Resources (DER), These include prosumer-owned small scale generation units, flexible loads, and storage devices. In a system with high penetration of renewable and distributed energy resources, supply-demand matching becomes a complicated task. To postpone or reduce excessive investments in infrastructure, innovative operational solutions are needed to make the demand follow the generation~\cite{ipakchi, IRENA}. Demand response is one of these solutions~\cite{peter}. It can be realised in several ways; while completely centralized architectures suffer from scalability problems~\cite{dr-transactive}, implementation of fully decentralized coordination approaches was found to be very challenging~\cite{centralizedvsdecentralized,p2p}. 

 Transactive control is a widely explored coordination architecture in which decisions are made by different actors based on economic signals. Market-based control is a sub-class of transactive control in which this is done through a market mechanism. Among such market-based approaches, iterative/negotiation-based approaches are challenged by poor convergence times, especially with huge prosumer participation as seen in~\cite{noncooperativeframework, peer2peer}. Real-time Market Based Control (RTMBC) combines the advantages of both centralized and decentralized methods of decision making~\cite{main}. Agents  submit their bid/offer functions based on local objectives and constraints. Allocation is determined via centralized market clearing in real-time~\cite{main}. However, these mechanisms perform poorly when used to coordinate uninterruptible loads. In~\cite{consensus}, oscillatory behaviour was observed when coordinating a population of Thermostatically Controlled Loads (TCLs). These challenges indicate the need for a coordination mechanism that is scalable, privacy-preserving, and usable by autonomous agents with low computational capabilities.

To address all these points of concern, Forecast Mediated Market Based Control (F-MBC) was developed in~\cite{main}, which aims to achieve cost-minimizing coordination by utilizing ``self-fulfilling forecasts''. The proposed scheme relies on a facilitator that broadcasts probabilistic price forecasts that correspond to a cost-optimal scenario over a scheduling horizon. Autonomous agents utilize the broadcast forecasts and a Markov Decision Process (MDP) based bidding policy to bid optimally, such that their own expected cost is minimized.  The scheme exhibited near-optimal system-level performance when used to coordinate a \emph{single population} of identical uninterruptible time-shiftable loads over multiple time-steps. Payments made by the agents were also found to be close to the optimal prices, thereby demonstrating favourable outcome from the device's perspective as well. 


This paper aims to investigate the coordination performance of F-MBC when used to schedule  \emph{heterogeneous populations} of deferrable loads across longer time scales. Specific contributions are:
\begin{itemize}
    \item Extending the facilitating agent described in~\cite{main} to generate reference prices for multiple populations of deferrable devices.
    \item Introducing a rolling-horizon based scheme to evaluate the performance through simulations.
    \item Developing a realistic scenario to gain insight about the scheme's performance, and analysing it in detail.
\end{itemize}

The organization of the paper is as follows:~\cref{Section: overview of MBC}  provides a brief context of the F-MBC mechanism. \cref{Section2} explains in detail the modifications made to the scheme to facilitate its usage for heterogeneous populations of devices. It also proposes a simulation methodology to test the robustness of the approach in near real-world settings. \cref{Section: results} describes a case study where the F-MBC mechanism is used to coordinate heterogeneous loads with dynamic power consumption cycles.

\section{Overview of the F-MBC scheme}
\label{Section: overview of MBC}
This section aims to illustrate the basic features of F-MBC scheme described in~\cite{main}, which will aid in better understanding of the extensions made to it. The main aim of F-MBC mechanism is to steer population(s) of self-dispatching flexible loads towards overall cost-minimization by utilizing ``self-fulfilling'' price forecasts, which are probabilistic in nature. It thus attempts to solve the optimal coordination problem: minimizing the overall cost of generation. It was shown in~\cite{main} that the solution to this problem corresponds to a Nash Equilibrium: devices purchasing energy at marginal cost would not derive any benefit from deviating from this globally optimal schedule.

\begin{figure}[ht]
	\centering
\includegraphics[width=0.7\linewidth,keepaspectratio]{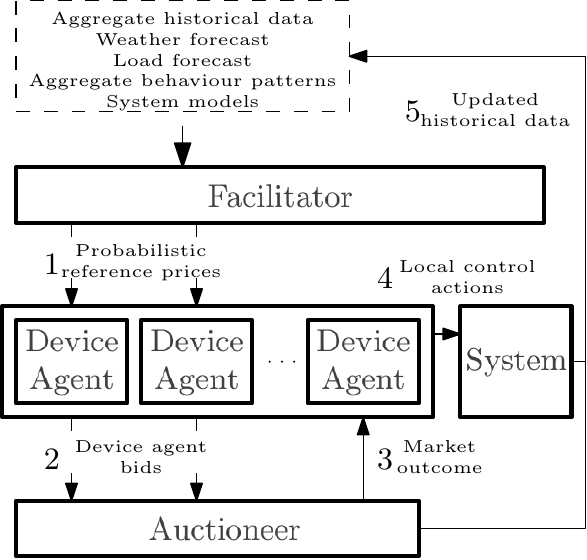}
	\caption{Pictorial representation of F-MBC approach~\cite{main}.}
	\label{F-MBC}
\end{figure}

The scheme is comprised of three different types of agents, namely the facilitator, the device agents and the auctioneer, as seen in~\cref{F-MBC}. These agents repeatedly execute the following process. At each time-step (with intervals $\Delta t=$15 minutes in the context of this paper), the facilitator provides price forecasts corresponding to a cost-optimal scenario by considering information such as weather forecast, predicted usage pattern of customers, system model, etc. 
The forecast prices are then communicated to device agents. Deferrable loads must run once, prior to their deadline, and cannot be interrupted. As a result, those that are not yet in the running state have two options: to start now or to wait. This choice is encoded in a device bid function.

Consider a device $j$ with a power demand profile $P_i^j$ for time-steps $i=0,\ldots,D^j-1$, where $D^j$ is the run duration. Under the assumption of independent probabilistic price forecasts $X_{t'}$ for $t'>t$, Theorem 1 of~\cite{main} describes an MDP-based optimal bidding policy for such a device agent. The bid function for a device with an \emph{optional start} at time $t$ is
\begin{subequations}
\begin{align}
 b_{t}^{j}(x)&=\left\{\begin{array}{ll}
P^{j}_{0} & x \leq \hat{x}_{t}^{j} \\
0 & x>\hat{x}_{t}^{j}
\end{array}\right. ,
\label{eq:bid function} \\
    \hat{x}_t^j&=\frac{C^{*j}_{t+1}-\sum_{i=1}^{D^j-1} \mathbb{E}[{X}_{t+i}] \cdot P_{i}^j \cdot \Delta t}{P_0^{j} \cdot \Delta t},
    \label{eq:xth}
    \end{align}
    \end{subequations}
where $C^{*j}_{t+1}$ is the expected optimal run cost if a decision is made to not start at this time-step (calculated by backward induction using Equation (8) of \cite{main}) and  ${E}[{X}_{t+i}]$ denotes the (finite) expectation of the price forecast at $t+i$.

The aggregate bid  and offer functions are then submitted to the auctioneer for market clearing. The aggregate bid function takes into account the inflexible load to be satisfied at time $t$, along with the bid functions from the flexible demand. The aggregate offer function represents both flexible and inflexible generation, in the form of a marginal cost function. It is assumed that the generating sources truthfully reveal their cost function to the auctioneer. The auctioneer then determines the clearing price, at the intersection of supply and demand curves. 
Agents execute local control actions based on the  clearing price $x_t$ and their previously submitted bids. For the bid function \eqref{eq:bid function}, device $j$ would start if the clearing price $x_t < \hat{x}_t^j$. If necessary, a tie-breaking procedure is used to differentiate between devices with identical bids~\cite{main}. 

This sequence of information exchange and device commitment occurs for every instant of the scheduling problem. F-MBC does not require the communication of private information (e.g.\ device characteristics, deadline, etc.\ ), to the auctioneer, thereby protecting end-user privacy. 

To test the internal consistency of the F-MBC scheme, a clairvoyant facilitator with complete information was used in~\cite{main} to generate reference prices. The prices corresponded to the outcome of a Mixed Integer Quadratic Program (MIQP) that determined the optimal number of device starts at each instant so that the overall generation cost is minimized. Noise was then added to these optimal reference prices to generate probabilistic reference prices.

\section{Facilitator generalizations}
\label{Section2}

The proof of principle of the F-MBC scheme in \cite{main} relied on a simplified setup where the optimal coordination problem was solved for a single population of constant-load devices, and for a single day. This paper relaxes both assumptions, necessitating extensions to the facilitator setup. We point out that we still consider a (near-)clairvoyant facilitator, in order to focus on the decentralised decision making by devices when given a near-optimal forecast. It is assumed that the deferrable loads only respond to the forecasts, and do not have access to any other price signals. 

\subsection{Heterogeneous loads}

First, we consider multiple populations $\mathcal{N}=\{1,\ldots, \nu \}$ of deferrable loads. Each population $n \in \mathcal{N}$ consists of identical, uninterruptible, time-shiftable devices with power profiles $P^n_i$ ($i=0,\ldots,D^n-1$). The aggregate flexibility of this population is characterised by the number of devices $\gamma_t^n$ becoming available to start at time $t$, and the number of devices $\phi_t^n$ having a deadline at time $t$, for all $t\in \mathcal{T}=\{1,\ldots,\tau\}$. 

To solve the optimal coordination problem it is required to determine the number of device starts $\sigma_t^n$ ( $\forall t \in \mathcal{T}; \forall n \in \mathcal{N}$) such that the overall costs, determined by the conventional generation level $P^{\text{g}}_t$, are minimized. This is achieved by solving the MIQP problem:
\begin{subequations}
\begin{flalign}
&       \underset{P_{t}^{\text{g}},\, \sigma_{t}^{n}}{\text{minimize}} 
   \sum_{t \in \mathcal{T}}\frac{1}{2} \frac{({P_{t}^{\text{g}}})^2}{k} \Delta t, & & \label{eq:obj}
\end{flalign}
subject to, $\forall t \in \mathcal{T}, n \in \mathcal{N}$:
\begin{align}
    P_{t}^{\text{g}} &\geq 0, \label{eq:pg}
\\
   P_{t}^{\text{g}}+P_{t}^{\text{r}} &\geq P_{t}^{\text{flex}}+P_{t}^{\text{l}}, \label{pb1}
\\
    P_{t}^{\text{flex}} &= \sum_{m \in \mathcal{N}}  P_{t}^{\text{flex},m}, \label{pb2} \\
P_{t}^{\text{flex},n} &= \sum_{i=0}^{\min(t-1,D^{n}-1)} \sigma_{t-i}^{n} P_{i}^{n},
\label{pb3} \\
 \sigma_{t}^{n} &\geq 0,  \label{eq:sigma0}
\\
      \sum_{i=1}^{t'}\sigma_{i}^{n} &\geq \phi_{t'+D^n}^{n},
\qquad  t' \in \{1,\ldots,\tau-D^{n}\} ,\label{dline1}
\\
  \sum_{i=1}^{t}\sigma_{i}^{n} &\leq \sum_{i=1}^{t}\gamma_{i}^{n}.
   \label{eq:o1}
\end{align}
The objective function~\eqref{eq:obj} is the integral of linear marginal costs of the conventional generating units where $P_{t}^{\text{g}}$ is the power generated at time $t$, $\,k \in \mathbb{R}_{+}$ is constant and $\Delta t$ is the fixed length of a time-step. Generator power limits and power balance constraints are expressed in ~\eqref{eq:pg} and ~\eqref{pb1} respectively. ~\cref{pb2} indicates that the total flexible demand at time $t$ is the sum of demands of all populations of flexible loads. $P_{t}^{\text{r}}$ is the available power generation from renewable sources (assumed to be without cost), while $P_{t}^{\text{flex}}$ and $P_{t}^{\text{l}}$ correspondingly represent the demand from flexible and inflexible loads at time $t$. \cref{pb3} expresses the power consumed by the devices belonging to a population $n$ at time $t$; where $D^{n}$ is the cycle duration of devices belonging to the population $n$, and $P_{i}^{n}$ is the power consumed by at the $i^{\text{th}}$ time-step of its cycle, by a device in population $n$. It also accounts for the uninterruptible nature of the devices. Equations \eqref{eq:sigma0}-\eqref{eq:o1} ensure that devices are scheduled in accordance with their availability and deadlines. 

Since F-MBC requires probabilistic price forecasts, uncertainty in forecasts must be simulated. To achieve this, we use the same modelling approach used in~\cite{main}.
However, the experiments in this paper do not aim at analyzing the effect of forecast uncertainty. Therefore, we only consider the case with a small maximum standard deviation of \SI{1}{\percent} of the average price.

\subsection {Simulating the facilitator in a realistic setting}

The problem \eqref{eq:obj}-\eqref{eq:o1} minimizes costs at once for a finite number of time-steps. However, in a real-world setting, the facilitator must serve updated price forecasts on an ongoing basis. We restate it as a rolling horizon problem, where self-fulfilling price forecasts are updated every time-step. 
This is represented graphically in~\cref{fig:rolling forecasts generation},  where forecasts are generated for \texttt{PH} future time-steps and control/scheduling is done every time-step by means of market clearing (\texttt{CH}=1). 
Note that in solving \eqref{eq:obj}-\eqref{eq:o1}, we assume to know the number of devices that will become available within the prediction horizon, and their respective deadlines.

\begin{figure}
  \centering
  {\includegraphics[width=\linewidth,keepaspectratio, scale=0.8]{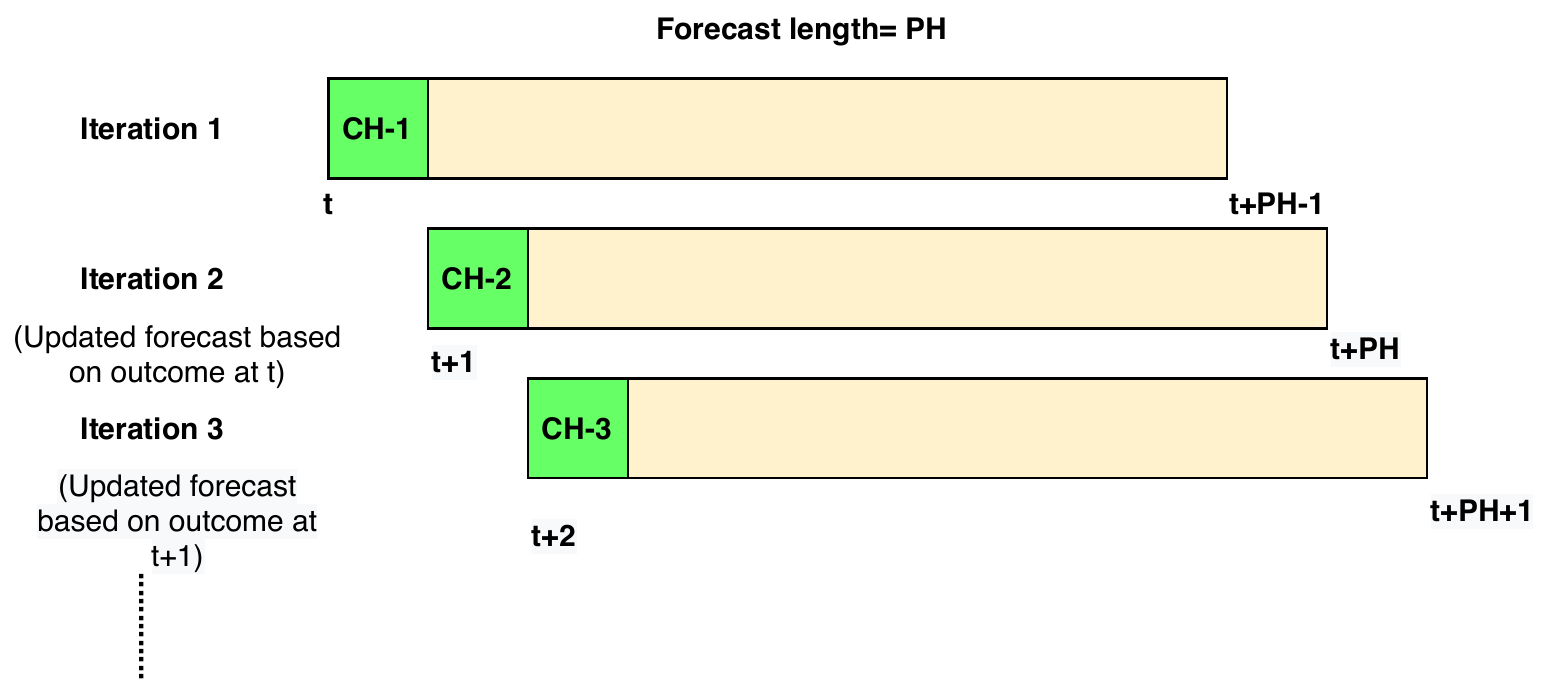}\label{fig:rh-reality}}
  \caption{Pictorial representation of rolling forecast generation during simulations. 
  }
 \label{fig:rolling forecasts generation}
\end{figure}

When using a finite prediction horizon \texttt{PH}, there is a possibility that boundary effects influence the control decisions within the scheduling horizon. In the case being considered, constraint \eqref{dline1} does not consider devices with a deadline later than $\tau$. effectively giving these devices a free pass to not run at all within the prediction horizon. As a result, problem \eqref{eq:obj}-\eqref{eq:o1} tends to schedule fewer devices than it would with a longer prediction horizon, leading to an underestimation of prices. For this reason, we shall refer to the resulting schedule as the \emph{optimistic forecast}. 

In order to gauge the robustness of the mechanism with respect to this prediction horizon, we also generate a \emph{pessimistic forecast}. This is created by adding the constraint
\begin{equation}
 \sum_{i=1}^{t}\sigma_{i}^{n} = \sum_{i=1}^{t}\gamma_{i}^{n},  \quad  t\in \{ \tau-D^{n}+1,\ldots, \tau \}, \forall n \in \mathcal{N}.
  \label{eq:pessimistic}
  \end{equation}
  \end{subequations}
This ensures that all devices which become available during the prediction horizon must start, and cannot make use of `free' energy at times $t>\tau$ unless absolutely necessary due to their late availability. It is expected that the pessimistic forecast leads to higher prices, especially for times near the prediction horizon, and therefore an increased eagerness for device agents to commence at the current time $t$.

\section{Experimental Analysis}
\label{Section: results}
This section describes the data used to develop the simulation scenario in which the performance of F-MBC was tested. The setup consists of a mix of renewable and flexible generation, which is used to satisfy both inflexible and flexible loads. These simulations were performed using MATLAB.

\subsection{Data generation}
\begin{figure}[ht]
    \centering
    \includegraphics[width=\linewidth,keepaspectratio]{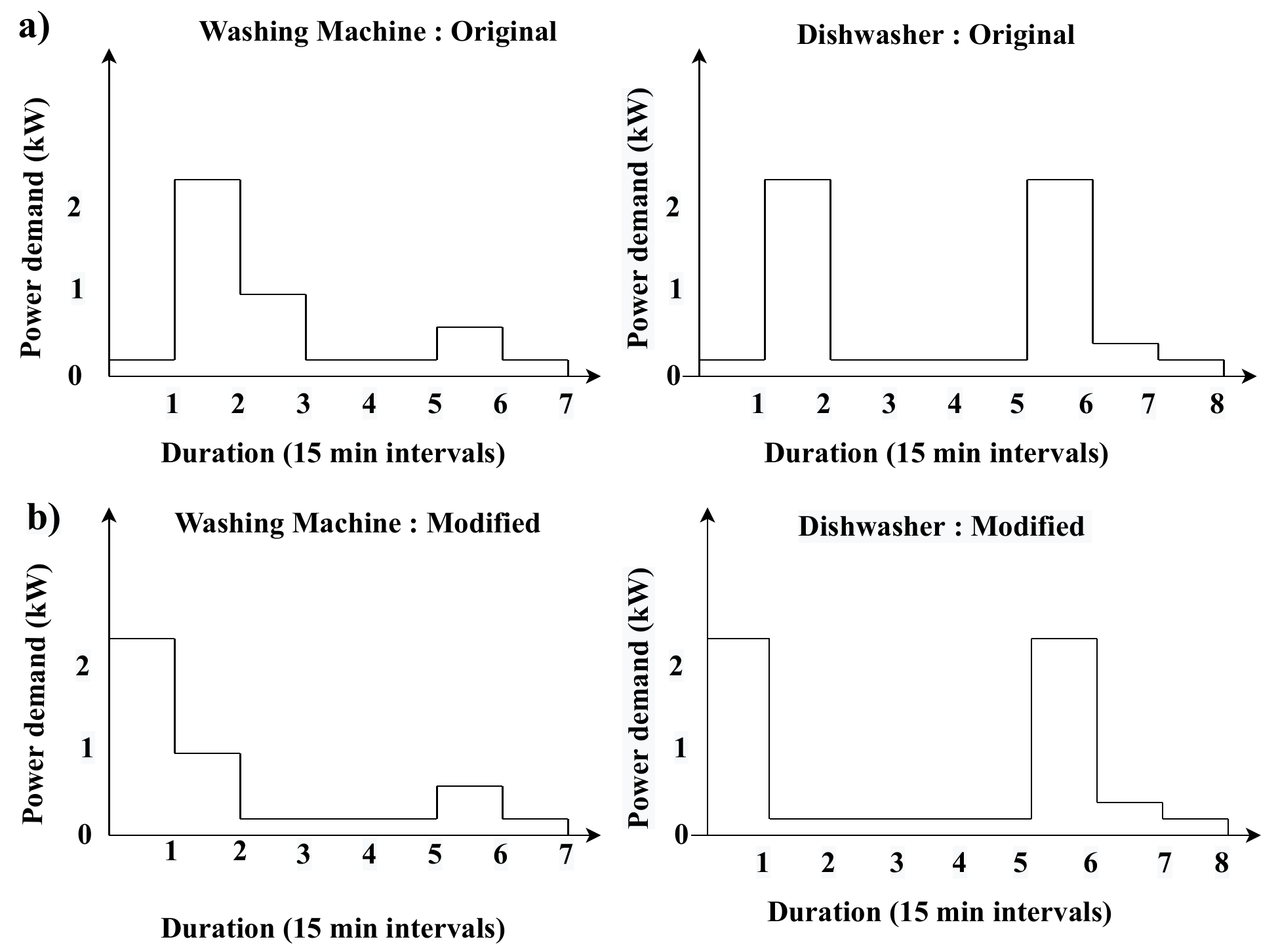}
    \caption{a) Load profiles of washing machines (left) and dishwashers (right), adapted from~\cite{stamminger2008}. b) Modified load profiles with high initial power demand.}
    \label{fig:load profile: rc}
\end{figure}

For the flexible loads, the load profiles of washing machines and dishwashers were considered. These devices have the highest penetration rates among residential devices in the Netherlands, of  \SI{94}{\percent} and \SI{58}{\percent} respectively~\cite{stamminger2008}. The load profiles of both device types are presented in~\cref{fig:load profile: rc}. These power consumption profiles are non-uniform, in contrast to those considered in~\cite{main}. Since the load profile have \SI{15} {\minute} intervals, the market is also assumed to clear every \SI{15}{\minute}. The bottom panels of~\cref{fig:load profile: rc} depict modified load profiles where the load peaks have been shifted to the first time-step, which results in markedly different performance of the distributed scheduling algorithm.

\begin{figure}[ht]
\centering
  {\includegraphics[width=\linewidth,keepaspectratio]{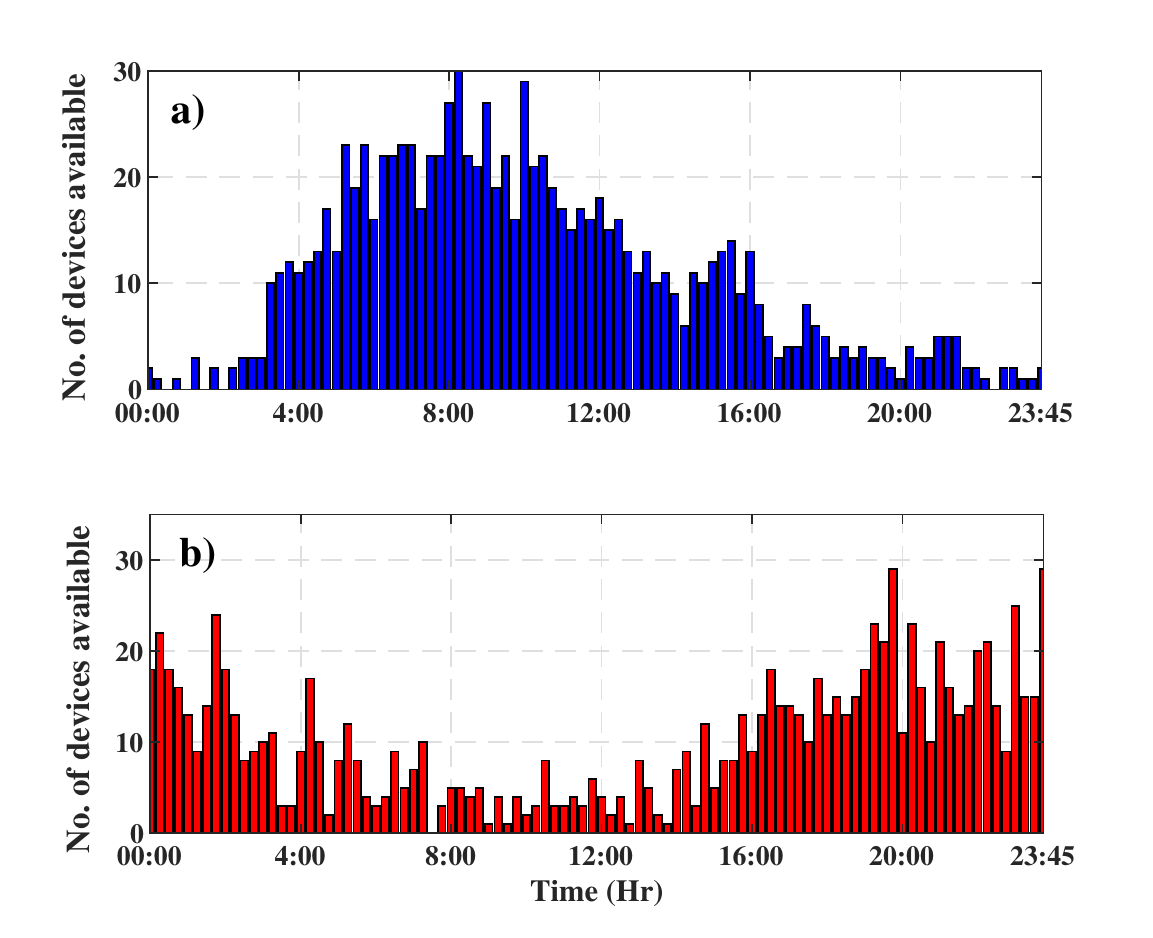}}
  \caption{a) Distribution of washing machine availabilities.
  \newline
  b) Distribution of dishwasher availabilities.}
 \label{fig:av}
\end{figure}

This experiment generates the availabilities of devices based on the average start times observed over the year 2013 in an experimental setup based in the cities of Utrecht and Amersfoort in the Netherlands~\cite{staats2017}. It is assumed that 1000 washing machines and diswashers become available on each day. From~\cite{staats2017}, it was evident that the preferable time for the operation of washing machines is in the morning hours. A visually similar profile was generated using a log-normal distribution ($\mu=0$, $\sigma=0.5$), scaled (from 00:00) to a median start time of 9:00 and rounded to the nearest \SI{15}{\minute} boundary. The availability trend of dishwashers is quite different, as the devices are available to start in large numbers from evening until early hours in the morning. For this, a log-normal distribution ($\mu=0$, $\sigma=0.25$), scaled to a median start time of 23:00 and wrapped around to the next day, was used. This led to the availability distribution of devices as shown in~\cref{fig:av}.

\begin{figure}[ht]
\centering
  {\includegraphics[width=\linewidth,keepaspectratio]{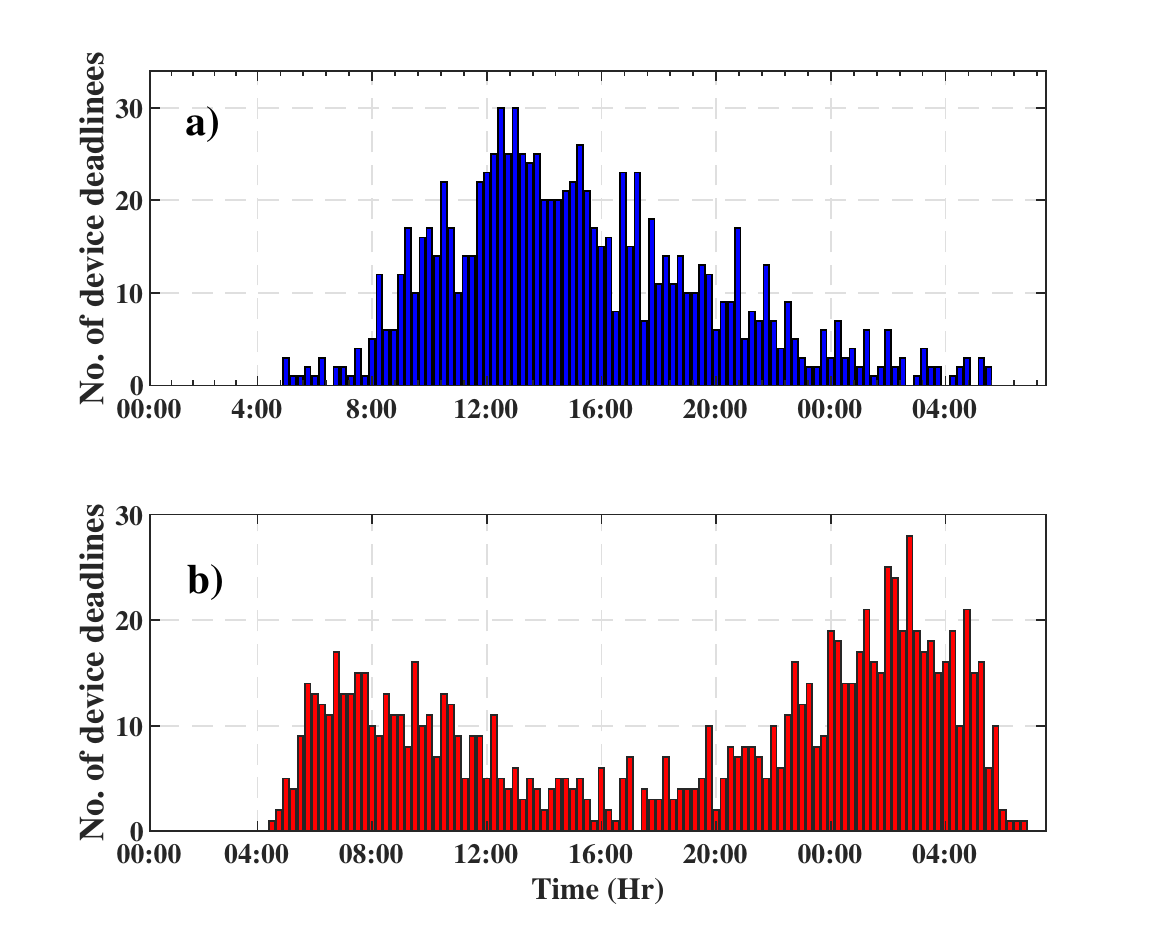}}
  \caption{a) Deadline distribution of washing machines
  \newline
  b) Deadline distribution of dishwashers}
 \label{fig:dline}
\end{figure}

To generate the deadlines, the findings from the SMART-A project are used in conjunction with the availability data. It was found in~\cite{stamminger2008} that the majority of the users who own devices with a delay function embedded in their devices prefer to defer their cycles for about \SI{3}{hours} from the time their devices are available to start. Therefore, deadlines are drawn from a normal distribution, centred on \SI{3}{hours} after initial availability, with a  standard deviation of \SI{0.5}{hours}. The deadline distribution of the devices in a day is given in~\cref{fig:dline}. It can be seen that devices that become available at the last few hours of the day have their deadlines in the early hours of the next day. The pattern of inflexible loads and renewable generation was taken to be identical to that in \cite{main}, averaged to \SI{15}{minutes} intervals and repeated for each simulated day.

\subsection{Simulation Parameters}
A period of 5 days was simulated, resulting in 480 time-steps.
At each instant, day ahead forecast prices are generated, with a prediction horizon of 96 intervals. 
The input data, namely the availabilities and deadlines of the washing machines and dishwashers, the inflexible load and generation are identical for each day, because the focus was on determining the `rolling' performance of the algorithm. Each simulation experiment was performed twice, using both optimistic and pessimistic scenarios to analyze the effect of the finite prediction horizon on the schedule.

\subsection{Results}

\begin{figure}[ht]
  \centering
   \includegraphics[width=\linewidth, keepaspectratio]{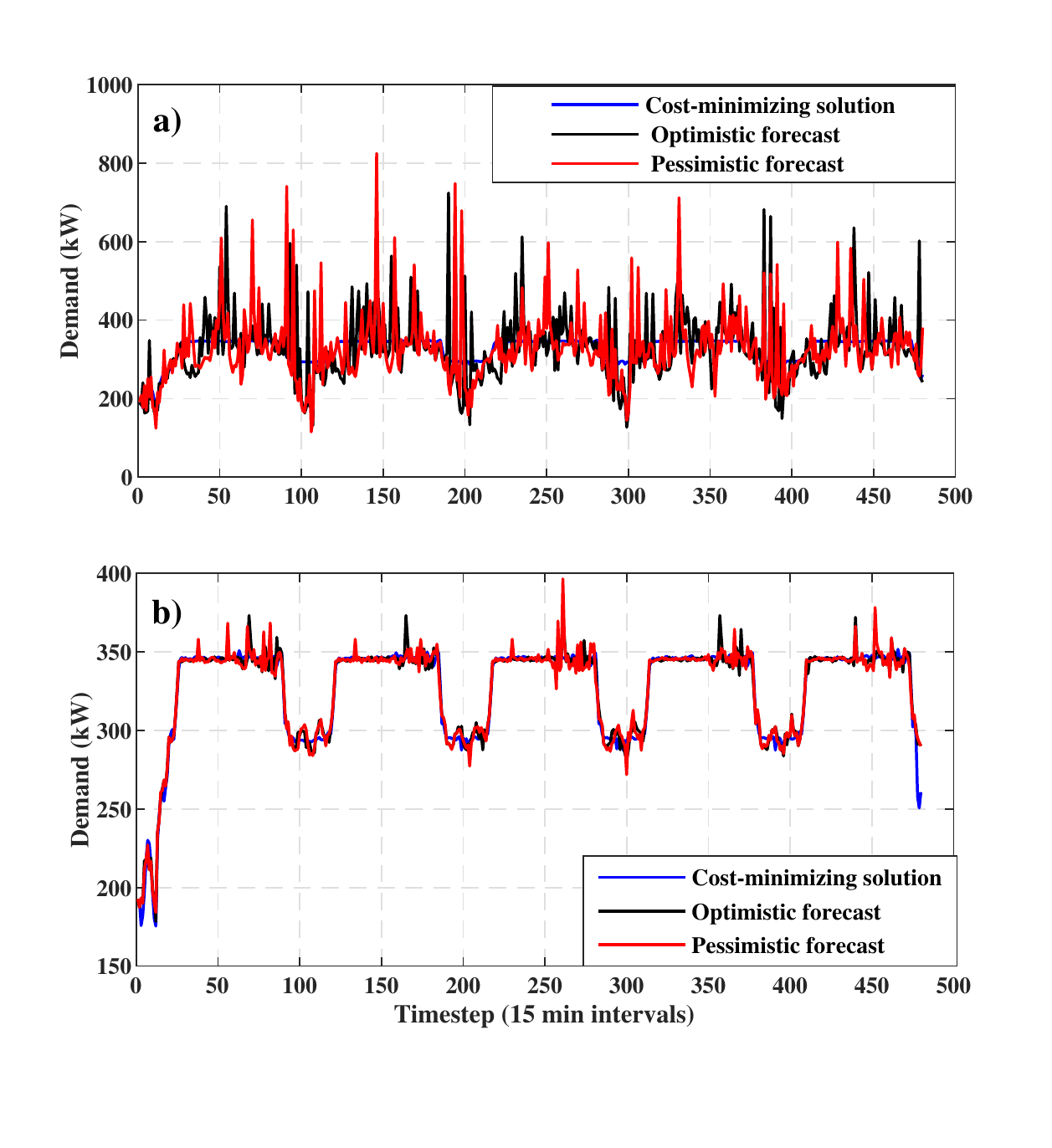}
  \caption{Total demand supplied by conventional generation. a) Coordination of devices with original load profiles. b) Coordination of devices with modified load profiles.}
 \label{fig:demand:rc}
\end{figure}

\cref{fig:demand:rc} shows the aggregate net demand profiles (supplied by the conventional generator) resulting from the use of the F-MBC mechanism. \cref{fig:demand:rc}a shows the results for the \emph{original device profiles} using optimistic (black line) and pessimistic forecasts (red line), alongside the optimal schedule obtained by solving \eqref{eq:obj}-\eqref{eq:o1} for the entire period (blue line). \cref{fig:demand:rc}b shows the scheduling result obtained for the appliances with \emph{modified device profiles}. One immediate observation is the large impact of device profiles on the resulting aggregate demand profile. In the following, we first analyse this performance difference in more depth, before identifying the underlying reasons.

Large deviations from the optimal schedule are visible for the original device profiles (\cref{fig:demand:rc}a), with the maximum and minimum power consumption being \SI{724}{\kilo \watt} and \SI{127}{\kilo \watt} respectively. The total system cost \eqref{eq:obj} achieved using F-MBC is \SI{6.3}{\percent} higher than the optimal value.

\begin{figure}[ht]
  \centering
   {\includegraphics[ width=\linewidth,keepaspectratio]{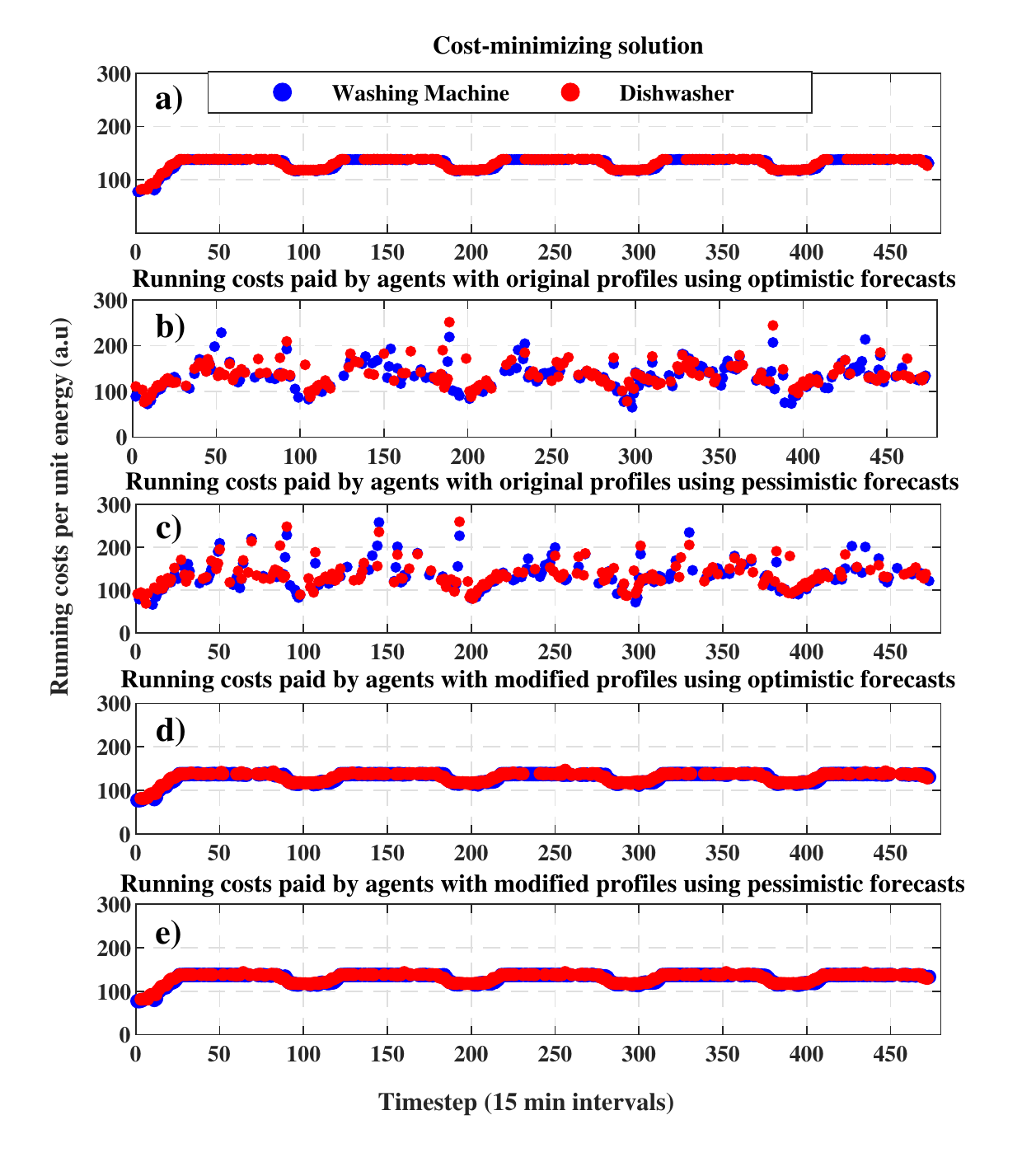}}
  \caption{Comparison of running costs for device agents in different scenarios, as a function of starting time. 
  }
 \label{fig:cs:rc}
\end{figure}

Further insight can be gained by considering the performance from the perspective of individual devices. \cref{fig:cs:rc} compares the total running cost for a device agent starting at time $t$ when coordinated with F-MBC and what agents would have paid under the cost-minimizing solution. Dots are only shown for those times at which at least one device starts. From \cref{fig:cs:rc}a, it can be seen that the agents pay almost a flat price, except at the start and the end of each day. The latter can be linked to the lower number of devices that become available and low inflexible net load at those instances. Recall that the cost-minimizing solution is a Nash equilibrium for this problem~\cite{main}.

Although the price paid by devices under the F-MBC mechanism is comparable, large price spikes are visible for devices with the \emph{original} demand profiles (\cref{fig:cs:rc}b/c). Costs were found to go as  high as \SI{228} (arbitrary units) and \SI{258} for washing machines and \SI{251} and \SI{259} for dishwashers using optimistic and pessimistic forecasts, respectively. On the other hand, for the case of \emph{modified} load profiles, the price paid by devices was very close to that obtained using a central optimization, indicating that the F-MBC mechanism was able to achieve near-optimal performance. 

Moreover, the results in Figures \ref{fig:demand:rc} and \ref{fig:cs:rc} suggest that the  choice between optimistic and pessimistic forecasts does affect the schedules, but not in a consistent manner. This suggests that the prediction horizon of 24 hours is sufficient in this case, and dispatch differences are due to rounding effects only. 

\begin{figure}[ht]
  \centering
  \subfloat[MIQP schedule ]{\includegraphics[width=0.5\linewidth]{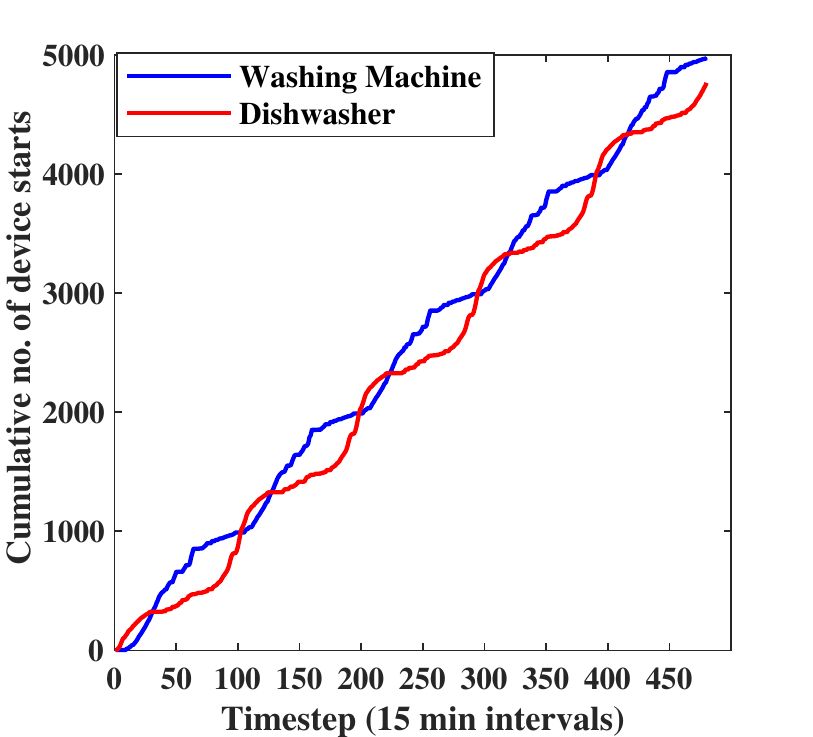}\label{fig:optimistic-MIQP}}
  \subfloat[F-MBC schedule ]{\includegraphics[width=0.5\linewidth]{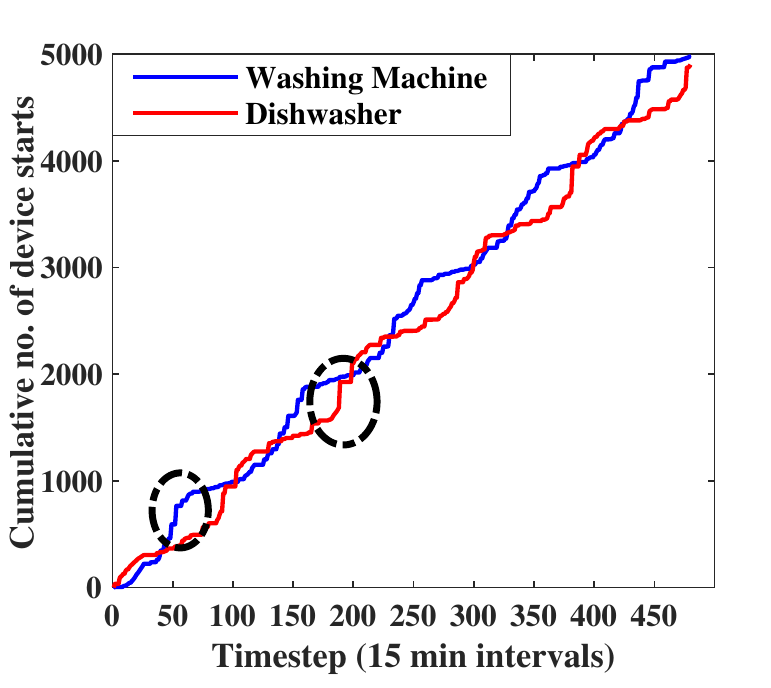}\label{fig:optimistic-FMBC}}
 \caption{Comparison of cumulative starts of devices. All results are for the original load profiles and optimistic price forecasts.}
 \label{fig:schedule comparison: optimistic:original}
\end{figure}

Finally, we investigate why the different device profiles in \cref{fig:load profile: rc} result in such large performance differences. The cumulative device start schedules for the \emph{original} device profiles, using optimal (MIQP) and F-MBC schedules, are compared in~\cref{fig:schedule comparison: optimistic:original}. It is clear that the F-MBC coordination process occasionally leads to bulk-starting events for both washing machines and dishwashers, evidenced by large steps in the start schedules. Such bulk-starting events are not visible in the case of the optimal schedule, nor when F-MBC is used with \emph{modified} device profiles (results not shown).

This observation suggests a reason for the observed performance difference between the two sets of device profiles. The original device profiles both have a 
very small initial power consumption (\SI{0.1}{\kilo \watt} and \SI{0.08}{\kilo \watt} respectively). This leads to a high variability of threshold bids, as the value $P^{j}_0$ features in the denominator of \cref{eq:bid function}. Moreover, each accepted bid only has a small effect on the supply curve at the time at which the device commits to starting. Small perceived cost differences in future steps can thus lead to large numbers of devices starting at the same time. Once allocated, these devices continue to run until they finish their cycle, due to the uninterruptible nature -- even if large price fluctuations occur due to the synchronised schedule of devices. 

This also explains why performance is greatly enhanced with the modified device profiles: they have their \emph{peak load} at the first time step, greatly reducing the sensitivity to rounding errors. It can therefore be concluded that the coordination performance of F-MBC mechanism is highly dependent on the power consumption profiles of the participating devices, and the number of such devices in the system.

\section{Conclusion}
This paper investigated the ability of F-MBC to coordinate heterogeneous populations of deferrable loads with dynamic power consumption patterns on a rolling basis. To do so, near-optimal price forecasts were generated by a facilitating agent, using a finite prediction horizon and optimistic or pessimistic constraints. 

The scheme was tested on two similar cases, differing only in the load sequence of the devices being scheduled. The F-MBC mechanism was found to result in near-optimal performance when devices had a high power demand at the beginning of their cycle. However, when devices had a low power consumption at the start of their cycle, performance was less satisfactory: high demand peaks, bulk switching and high agent payments were observed. 

It is likely that this performance gap is reduced for larger systems with more deferrable loads, or systems with a greater diversity of such loads. Further theoretical and simulation studies are required to quantify this performance gap, and to propose suitable adjustments to the scheme. Other relevant extensions to the F-MBC scheme are applications to other device types such as interruptible loads and storage devices, or the use of dependent forecast errors in bid formation.

    \FloatBarrier
\bibliographystyle{IEEEtran}


\end{document}